\documentclass{jps-cp}
\usepackage{txfonts} %Please comment out this line unless the txfonts package is availabe in your LaTeX system.
\usepackage{bm}

\title{Topological Superconducting Phases in a Strained Altermagnet–Superconductor Heterostructure}

\author{Keita Yoshizawa, Ryo Okugawa and Takami Tohyama}

\inst{Department of Applied Physics, Tokyo University of Science, Katsushika, Tokyo 125-8585, Japan}

\abst{
We investigate topological superconductivity in a heterostructure consisting of a two-dimensional $s$-wave superconductor and a $d$-wave altermagnet with Rashba spin-orbit coupling.
In particular, we study the effects of strain and hopping anisotropy on the topological superconducting phases.
By calculating the Chern number,
we obtain topological phase diagrams as functions of the chemical potential and the strength of the effective magnetic field induced by the magnetic proximity effect. 
We find that strain-induced lattice distortion allows topological superconducting phases to emerge over a broad parameter region. 
Our findings suggest that strain that lowers symmetry can serve as a route to realizing topological superconductivity in altermagnet–superconductor heterostructures.
}

\kword{Altermagnet, Topological Superconductor, Rashba Spin-Orbit Coupling}

\begin{document}
\maketitle

\section{Introduction}

Topological superconductors have been intensively studied as a platform for fault-tolerant quantum computation based on Majorana bound states \cite{Alicea2012,Nayak2008,Sato2017}.
In particular, an \(s\)-wave superconductor with Rashba spin-orbit coupling, coupled to a ferromagnet, has been considered a promising system for realizing topological superconductivity.
In two-dimensional heterostructures consisting of an \(s\)-wave superconductor and a ferromagnet, 
topological superconducting phases emerge over broad ranges of the chemical potential and the strength of an effective magnetic field \cite{Sato2010,Sato2009}.  

Recently, it has been theoretically proposed that Majorana edge modes can be realized in a heterostructure consisting of an $s$-wave superconductor and a $d$-wave altermagnet, providing an alternative to conventional ferromagnet-based systems \cite{Hadjipaschalis2025,Ghorashi2024,Liudeng2026,Zhu2026,Alam2026}.
Altermagnetism is a new class of magnetic order distinct from ferromagnetism and antiferromagnetism \cite{Šmejkal2020,Naka2019,Šmejkal2022Beyond,Šmejkal2022Emerging}. 
Altermagnets have antiparallel spin arrangements and zero net magnetization similar to conventional antiferromagnets, while they exhibit spin-split electronic bands due to time-reversal symmetry breaking.
Indeed, recent theoretical studies have demonstrated the altermagnetic proximity effect \cite{Zhu2026,Heinsdorf2026},
which supports the realization of topological superconductivity without relying on ferromagnets.
Using the proximity effect, a heterostructure that consists of V$_2$Se$_2$O and an $s$-wave superconductor has been suggested as a platform for topological superconductivity \cite{Zhu2026, Liudeng2026}.
In the altermagnetic counterpart, previous studies have reported that anisotropic hopping in the superconducting layer plays a significant role in topological superconducting phases with finite Chern numbers  \cite{Zhu2026,Ghorashi2024,Alam2026}.
This observation suggests that strain may promote topological superconductivity by lowering the crystalline symmetry.
However, its effects on topological superconductivity in altermagnet-superconductor heterostructures remain largely unexplored.

\begin{figure}[tb]
\centering
\includegraphics[width=10cm]{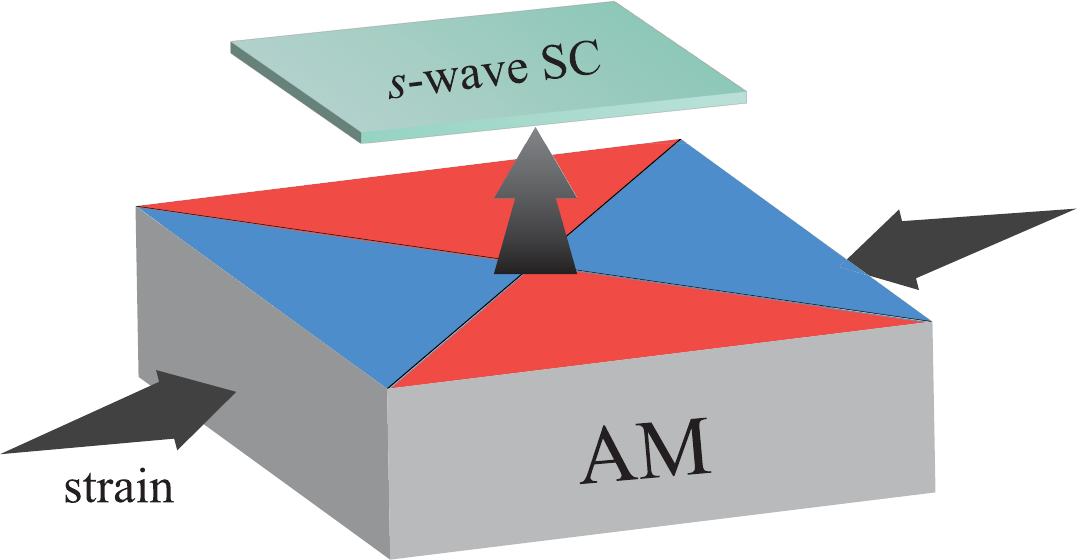}
\caption{Schematic of a heterostructure consisting of a two-dimensional \(s\)-wave superconducting layer (SC) and a \(d\)-wave altermagnet (AM).
}
\label{f1}
\end{figure}

In this work, we investigate the effects of strain and hopping anisotropy on topological superconductivity in a heterostructure consisting of an \(s\)-wave superconductor with Rashba spin-orbit coupling and a \(d\)-wave altermagnet [Fig.~\ref{f1}].
We determine topological phase diagrams as functions of the chemical potential and the proximity-induced effective magnetic field by calculating the Chern number.
We show that strain enables topological superconducting phases with nonzero Chern numbers even in the absence of hopping anisotropy.
Furthermore, when strain and hopping anisotropy coexist, topological superconducting phases emerge over a broader parameter range.
We also demonstrate the existence of Majorana edge modes in the topological superconducting phases.

\section{Model}
We study a model of a heterostructure consisting of a square-lattice $s$-wave superconductor and a $d$-wave altermagnet under strain.
The Bogoliubov–de Gennes  Hamiltonian \(\mathcal{H}_{\mathrm{BdG}}(\bm{k})\) in the Nambu basis \(\Psi_{\bm{k}} =(c_{\bm{k},\uparrow},c_{\bm{k},\downarrow},c^\dagger_{-
\bm{k},\uparrow},c^\dagger_{-\bm{k},\downarrow})^T\) is given by
\begin{equation}
\mathcal{H}_{\mathrm{BdG}}(\bm{k}) =
    \begin{pmatrix}
    \mathcal{H}(\bm{k}) & i\Delta _s\sigma _y \\
    -i\Delta _s^*\sigma _y & -\mathcal{H}^T(-\bm{k})
    \end{pmatrix}\ ,
\label{e3}
\end{equation}
where the normal-state Hamiltonian $\mathcal{H}(\bm{k})$ is given by
\begin{equation}
\mathcal{H}(\bm{k})=
(-2t_x\cos k_x -2t_y\cos k_y -\mu)\sigma_0
+M(\bm{k}) \sigma_z\ 
+\alpha(\sin k_y\sigma_x -\sin k_x\sigma_y).
\label{e1}
\end{equation}
Here, \(t_x\) and \(t_y\) are the hopping amplitudes in the \(x\) and \(y\) directions, respectively, \(\mu\) is the chemical potential,
and $\Delta _s$ is the superconducting order parameter. 
We set lattice constants to unity.
In Eqs.~(\ref{e3}) and (\ref{e1}),
\(\sigma_0\) denotes the \(2\times2\) identity matrix, and \(\sigma_i\ (i=x,y,z)\) are the Pauli matrices in spin space. 
The second term in Eq.~(\ref{e1}) describes an effective magnetic field induced by the magnetic proximity effect from the altermagnet, where
\begin{align}
    M(\bm{k})=J_{\mathrm{AM}}(\cos k_x -b\cos k_y).
\end{align}
$J_{\mathrm{AM}}$ denotes the strength of the effective field and
$b$ is a dimensionless parameter that characterizes strain-induced lattice distortion in the altermagnet.
Such an effect can emerge when the crystalline symmetry is lowered \cite{Ghorashi2024,Karetta2025,Naka2025}.
The third term represents the Rashba spin-orbit coupling.
This model for $b=1$ has been investigated as a platform for topological superconductivity \cite{Zhu2023,Zhu2026,Ghorashi2024}.
Throughout this paper, we fix \(\alpha=0.4t_x\) and \(\Delta _s=0.2t_x\)
and assume $b>0$.

Because topological phase transitions are accompanied by gap closings,
we calculate the energy eigenvalues of the model.
The eigenvalues of $\mathcal{H}_{\mathrm{BdG}}(\bm{k})$ are given by $\pm E_{\pm}(\bm{k})$, where
\begin{equation}  
E_{\pm}(\bm{k})=
\left[
\varepsilon (\bm{k} )^2  + M(\bm{k})^2 + R(\bm{k})^2 + |\Delta _s|^2
\pm 2\sqrt{\varepsilon (\bm{k})^2 M(\bm{k})^2+  R(\bm{k})^2\varepsilon (\bm{k})^2+|\Delta _s|^2 M(\bm{k} )^2}
\right] ^{1/2},
\end{equation}
with 
$\varepsilon (\bm{k})=-2t_x\cos k_x -2t_y\cos k_y -\mu$
and
$R(\bm{k})=|\alpha| \sqrt{\sin ^2k_x + \sin ^2 k_y}$.
The quasiparticle band gap closes when ${E}_{-}(\bm{k})=0$.
After a straightforward calculation, we can rewrite this condition as 
\begin{align}
    \varepsilon (\bm{k}) ^2 + |\Delta _s|^2 = M(\bm{k})^2, \hspace{3mm}
    \sin k_x = \sin k _y =0.
\end{align}
Therefore, the gap can close only at the high-symmetry points $\mathrm{\Gamma}=(0,0), \mathrm{X}=(\pi,0), \mathrm{Y}=(0,\pi)$, and $\mathrm{M}=(\pi, \pi)$ 
if $M(\bm{k})\neq 0$ at these points.
In addition, when $t_x=t_y$, the energy gap closes simultaneously at the X and Y points.

\begin{figure}[tb]
\centering
\includegraphics[width=15cm]{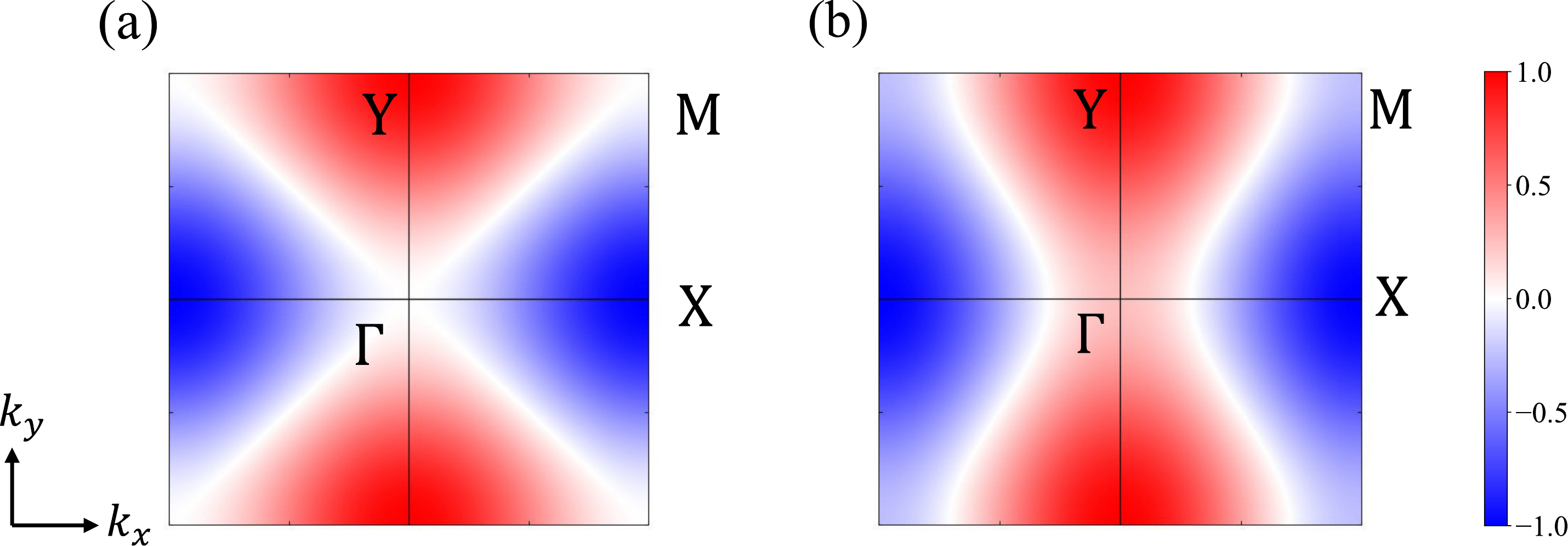}
\caption{Momentum dependence of the normalized effective magnetic field $\tilde{M}(\bm{k})$ for (a) \(b=1\) and (b) \(b=0.6\). }
\label{f2}
\end{figure}

As seen from the above calculations, 
the effective magnetic field $M(\bm{k})$ plays an important role in gap closing.
Figure \ref{f2} displays the momentum dependence of the normalized effective magnetic field defined as
$\tilde{M} (\bm{k}) = {M(\bm{k})}/{[J_{\mathrm{AM}}(1+b)]}$
for $b=1$ and $b\neq 1$.
In the absence of the distortion $(b=1)$,
the field vanishes at the \(\Gamma\) and \(\mathrm{M}\) points [Fig.~\ref{f2}(a)].
In contrast, for $b\neq 1$, the effective magnetic field is finite at all the high-symmetry points, as shown in Fig.~\ref{f2}(b).
Therefore, we expect that the strain-induced distortion leads to a richer topological phase diagram because the gap closings can occur at all the high-symmetry points.

\section{Topological Phase Diagrams}
Here, we study topological superconductivity characterized by a nonzero Chern number.
To obtain topological phase diagrams, we compute the Chern number defined as
\begin{equation}
C=\frac{i}{2\pi}\int_{\mathrm{BZ}}d^2k\,
\mathrm{Tr}F(\bm{k})
\label{e4}
\end{equation}
where $F(\bm{k})$ is the non-Abelian Berry curvature of the negative-energy eigenstates.
The integration is performed over the two-dimensional Brillouin zone.
To evaluate the Chern number numerically, we use the Fukui--Hatsugai--Suzuki method \cite{Fukui2005}.
In the present model, we assume that the parameter \(b\) and the hopping anisotropy can be varied independently, since they characterize properties of different layers.

\begin{figure}[tb]
\centering
\includegraphics[width=13cm]{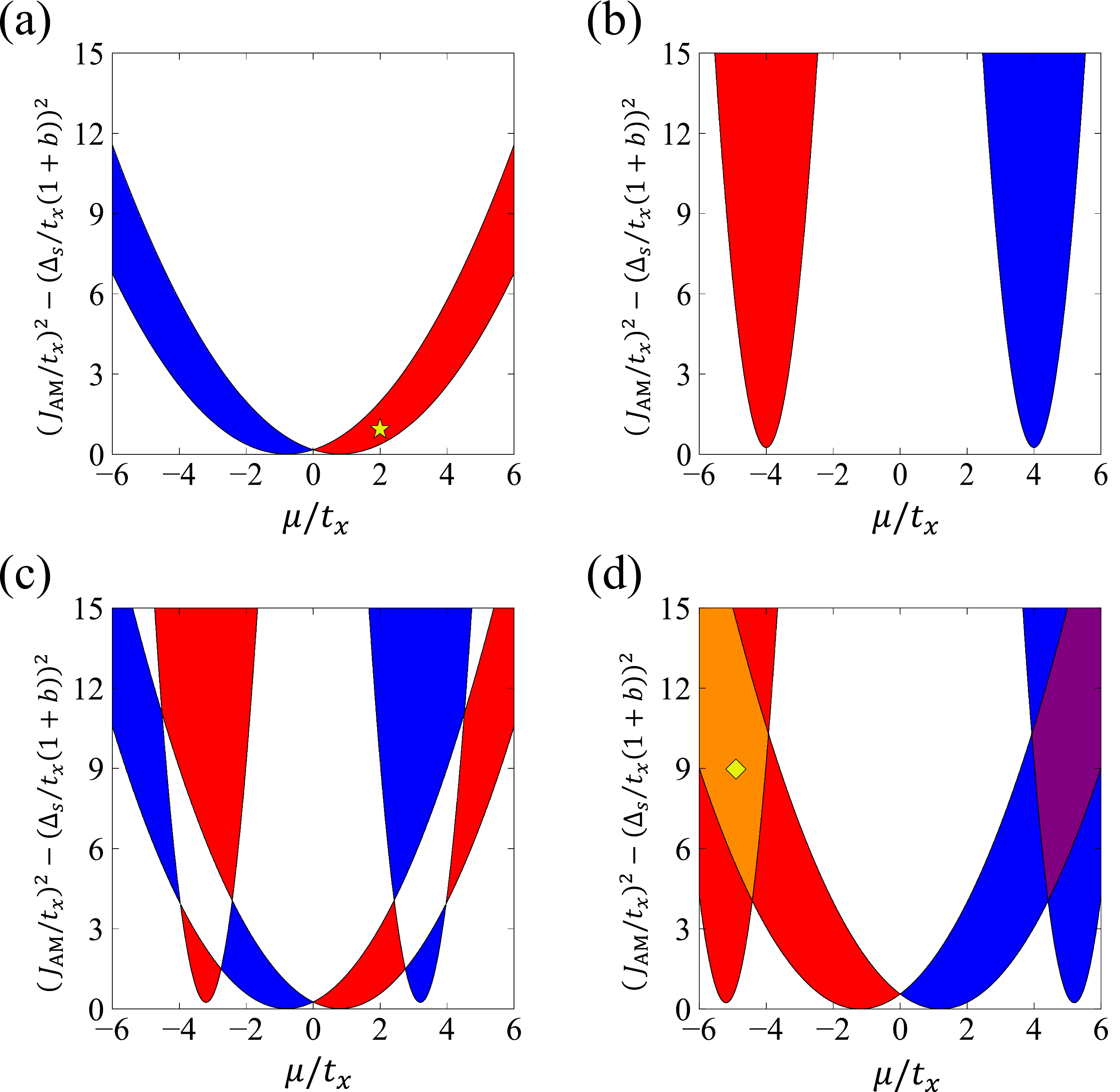}
\caption{Topological phase diagrams of the heterostructure for (a) \(b=1\) and \(t_y=0.6t_x\), (b)
\(b=0.6\) and \(t_x=t_y\), (c) \(b=0.6\) and \(t_y=0.6t_x\), and (d) \(b=0.6\) and \(t_y=1.6t_x\).
The red, blue, orange, and purple regions indicate \(C=1\), \(C=-1\), \(C=2\), and \(C=-2\), respectively. The star in (a) and the diamond in (d) indicate the parameter sets used to calculate the edge spectra in Fig.~\ref{f4}.}
\label{f3}
\end{figure}

For comparison with the case with distortion ($b\neq 1$), 
we first examine the system with $b=1$ and anisotropic hopping $t_x\neq t_y$.
The resulting topological phase diagram is shown in Fig.~\ref{f3}(a).
The phase diagram shows the emergence of topological superconducting phases with $|C|=1$.
The appearance of the topological phase with $|C|=1$ is consistent with the results of previous works \cite{Zhu2026,Ghorashi2024}.

We now discuss the topological phase transitions.
As the strength of the effective magnetic field $J_{\mathrm{AM}}$ is increased, gap closings occur at the X and Y points, leading to topological phase transitions.
As shown in Fig.~\ref{f2}(a),
since the effective magnetic field has opposite signs at the X and Y points, 
gap closings at the two points lead to changes in the Chern number with opposite signs.
Each gap closing changes the Chern number by $\pm 1$, and therefore the maximum value of $|C|$ is one as $J_{\mathrm{AM}}$ is increased.
For $t_x=t_y$, a nonzero Chern number is not allowed in this model with $b=1$
since the energy gap closes simultaneously at the X and Y points. 

Hereafter, we investigate the effect of lattice distortion, which is characterized by $b\neq 1$.
We start with the case of isotropic hopping for $t_x=t_y$.
Figure~\ref{f3}(b) shows the phase diagram for \(b=0.6\).
We find that topological superconducting phases emerge even when the hopping is isotropic.  
The emergence of these phases can be understood from the momentum dependence of the effective magnetic field $M(\bm{k})$.
For \(b\neq1\), the effective magnetic field is nonzero at the points \(\Gamma\) and \(\mathrm{M}\), as shown in Fig.~\ref{f2}(b).
Therefore, the bulk energy gap can close at these points, leading to topological phase transitions. 

Moreover, we consider the system with both lattice distortion and anisotropic hopping.
We present topological phase diagrams for $t_y=0.6t_x$ and $t_y=1.6t_x$ in Figs.~\ref{f3}(c) and \ref{f3}(d), respectively.
These phase diagrams show topological superconducting phases over a broader parameter region.
In this case, the energy gap can close at all the high-symmetry points $\Gamma$, X, Y, and M, and the gap closings do not occur simultaneously.
At each gap closing, the Chern number changes by $\pm 1$. 
Consequently, topological superconducting phases with $|C|=2$ can emerge as well as those with $|C|=1$, depending on the order of the gap closings as $J_{\mathrm{AM}}$ is increased.
Therefore, a variety of topological phase transitions can occur when $b \neq 1$ and $t_x\neq t_y$.

\begin{figure}[t]
\centering
\includegraphics[width=15cm]{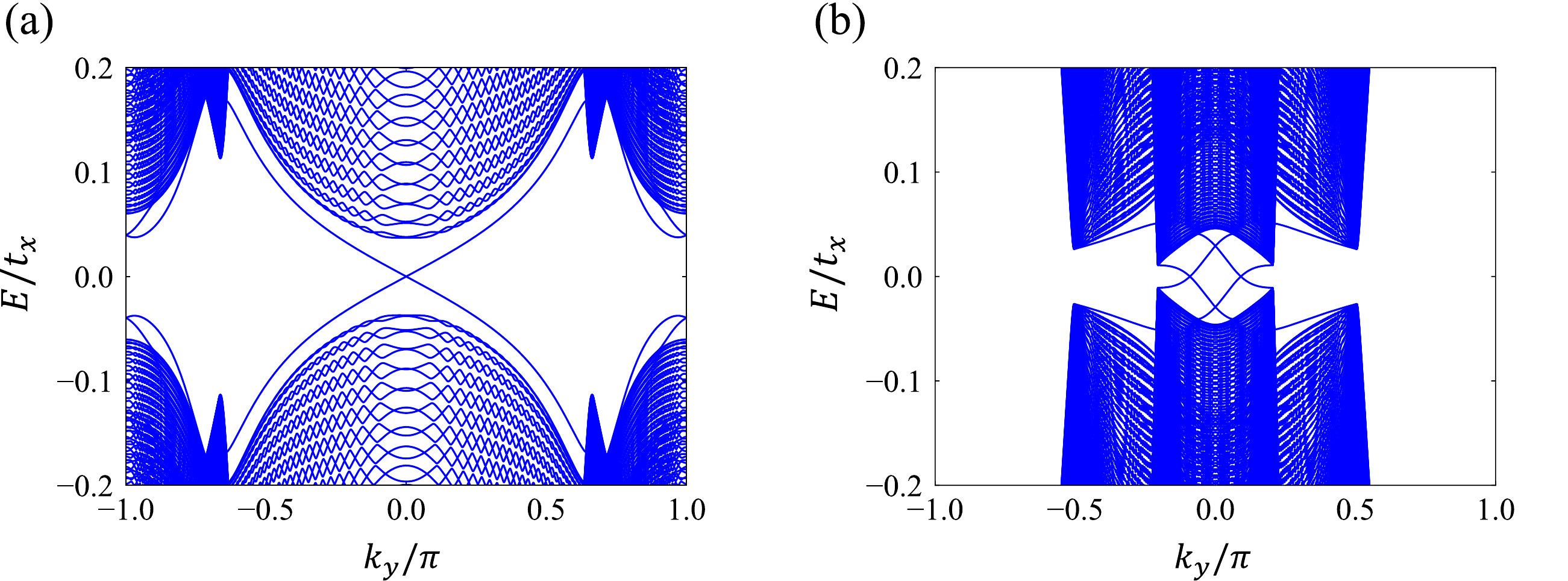}
\caption{Energy spectra for the parameter sets indicated by (a) the star in Fig.~\ref{f3}(a) and (b) the diamond in Fig.~\ref{f3}(d).
The star and diamond correspond to $\mu =2t_x$ and $J_{\mathrm{AM}}=t_x$, and  $\mu=-5t_x$ and $J_{\mathrm{AM}}=3t_x$, respectively.
}
\label{f4}
\end{figure}

Finally, we confirm the existence of Majorana edge states in the topological superconducting phases as a manifestation of the nontrivial bulk topology.
To this end, we impose open boundary conditions in the \(x\) direction and periodic boundary conditions in the \(y\) direction. 
Figures~\ref{f4}(a) and \ref{f4}(b) show the energy spectra under these boundary conditions for the parameter sets marked by the star and diamond in Fig.~\ref{f3}, respectively. 
We find that gapless Majorana edge modes appear within the bulk energy gap.
The number of gapless edge modes on a single edge is consistent with the absolute value of the Chern number, \(|C|\).
These results demonstrate that Majorana edge states can be realized using altermagnets.

\section{Conclusion}
We have investigated the effects of strain and hopping anisotropy on topological superconductivity in an altermagnet–superconductor heterostructure. 
Because symmetry lowering allows gap closings to occur at all the high-symmetry points, topological superconducting phases with Chern numbers $C=\pm1$ and $\pm2$ can emerge.
We have also confirmed the emergence of gapless Majorana edge modes in the topological superconducting phases.
Our results suggest that strain provides an effective means of realizing and  controlling topological superconductivity in altermagnet–superconductor heterostructures.

\section{Acknowledgment}
We thank S. Sumita for valuable comments. 
This work was supported by JSPS KAKENHI (Grants No. JP24K00586 and JP25H01248).

\end{document}